\documentclass[twocolumn, journal]{IEEEtran}

\newcommand{\etal}{\emph{et al.} }

\usepackage{color}
\usepackage{multirow}

\usepackage[pdftex]{graphicx}

\usepackage{amssymb}
\usepackage{amsmath}

\begin{document}

\title{A Blockchain-based Decentralized Federated Learning Framework with Committee Consensus}


\author{\IEEEauthorblockN{Yuzheng~Li, Chuan~Chen\IEEEauthorrefmark{1}, Nan~Liu, Huawei~Huang, Zibin~Zheng and Qiang~Yan\IEEEauthorrefmark{2}}

\IEEEauthorblockA{School of Data and Computer Science, Sun Yat-sen University, Guangzhou, China}

\IEEEauthorblockA{National Engineering Research Center of Digital Life, Sun Yat-sen University, Guangzhou, China}

\IEEEauthorblockA{\IEEEauthorrefmark{2}WeBank Co. Ltd., China}

\IEEEauthorblockA{\IEEEauthorrefmark{1}Email: chenchuan@mail.sysu.edu.cn}
}

\maketitle

\begin{abstract}
Federated learning has been widely studied and applied to various scenarios. In mobile computing scenarios, federated learning protects users from exposing their private data, while cooperatively training the global model for a variety of real-world applications. However, the security of federated learning is increasingly being questioned, due to the malicious clients or central servers' constant attack to the global model or user privacy data. To address these security issues, we proposed a decentralized federated learning framework based on blockchain, i.e., a Blockchain-based Federated Learning framework with Committee consensus (BFLC). The framework uses blockchain for the global model storage and the local model update exchange. To enable the proposed BFLC, we also devised an innovative committee consensus mechanism, which can effectively reduce the amount of consensus computing and reduce malicious attacks. We then discussed the scalability of BFLC, including theoretical security, storage optimization, and incentives. Finally, we performed experiments using real-world datasets to verify the effectiveness of the BFLC framework.
\end{abstract}

\begin{IEEEkeywords}
Blockchain, Federated learning, Validation consensus, Robustness
\end{IEEEkeywords}



\section{Introduction}
With the introduction of GDPR, both industry and academia began to pay more attention to the privacy protection of machine learning. User-generated private data should not be exposed or uploaded to a central server. Google proposed Federated Learning (FL) in 2016 to solve the problem of collaborative training for privacy protection. FL proposes a distributed training model with two roles: the participating devices and the central server. Instead of uploading private data, nodes locally update the global model and then upload the model updates (i.e., the local gradients). The central server collects these updates and integrates them to form an updated model. Because of this privacy feature, FL is attracting more and more researchers' attention in recent years.

In FL settings, a server performs the central operations of update aggregation, client selection, global model maintenance. The server needs to collect updates from numerous clients to complete the aggregation operation, and it also needs to broadcast a new global model to these clients, which puts a high demand on network bandwidth. Also, cloud-based servers are affected by the stability of cloud service providers \cite{DBLP:journals/corr/KonecnyMYRSB16}. A centralized server can skew the global model by favoring some clients. Moreover, some malicious central servers can poison the model 
and even collect clients' privacy data 
from updates. Therefore, the stability, fairness, and security of the central server are crucial to FL.

A direct idea is to remove the server and execute the central tasks on distributed client nodes. The blockchain, which is viewed as decentralized storage, can serve as the basis for maintaining FL. In detail, we can design protocols to execute the aggregation task on clients. BAFFLE \cite{DBLP:journals/corr/abs-1909-07452} mentions using blockchain to store and share the global model, and using perform aggregation with smart contracts. With the removal of the central server, the above challenges need not be considered. However, the computation and network transmission pressure of this task are all transferred to the nodes. In particular, when all nodes have to deal with consensus tasks, the computational overhead per round is huge.

Zhou \etal \cite{DBLP:journals/corr/abs-1912-07860} propose using blockchain to maintain the global model within a community and reach a consensus, and leveraging allreduce protocol
to transmit and update the model among multiple communities. The global model is updated continuously and promoted by various communities. Chen \etal \cite{8622598} propose to leverage the blockchain to record the updates from nodes and the evaluation of that updates. Underrated nodes may be kicked out of the community as a defense against malicious devices. However, maintaining multiple blockchains at the same time \cite{DBLP:journals/corr/abs-1912-07860} is not conducive to model sharing, and nodes in different communities can hardly obtain models or update records of other communities. If a community as a whole is malicious, it is difficult for other honest communities to detect and resist, then a trusty global detection might be needed
.

Through the literature review, it would be an effective way for blockchain to serve as an effective decentralized storage and replace the central FL servers. However, the efficiency of consensus is in urgent need of improvement. Although storing models and model updates in blockchain brings many security advantages, it is also a huge burden of the storage capacity on blockchain nodes. Therefore, how to reduce the consumption of a blockchain-based FL is also a key challenge.

\begin{figure*}[t]
    \centering
    \includegraphics[width=.9\linewidth]{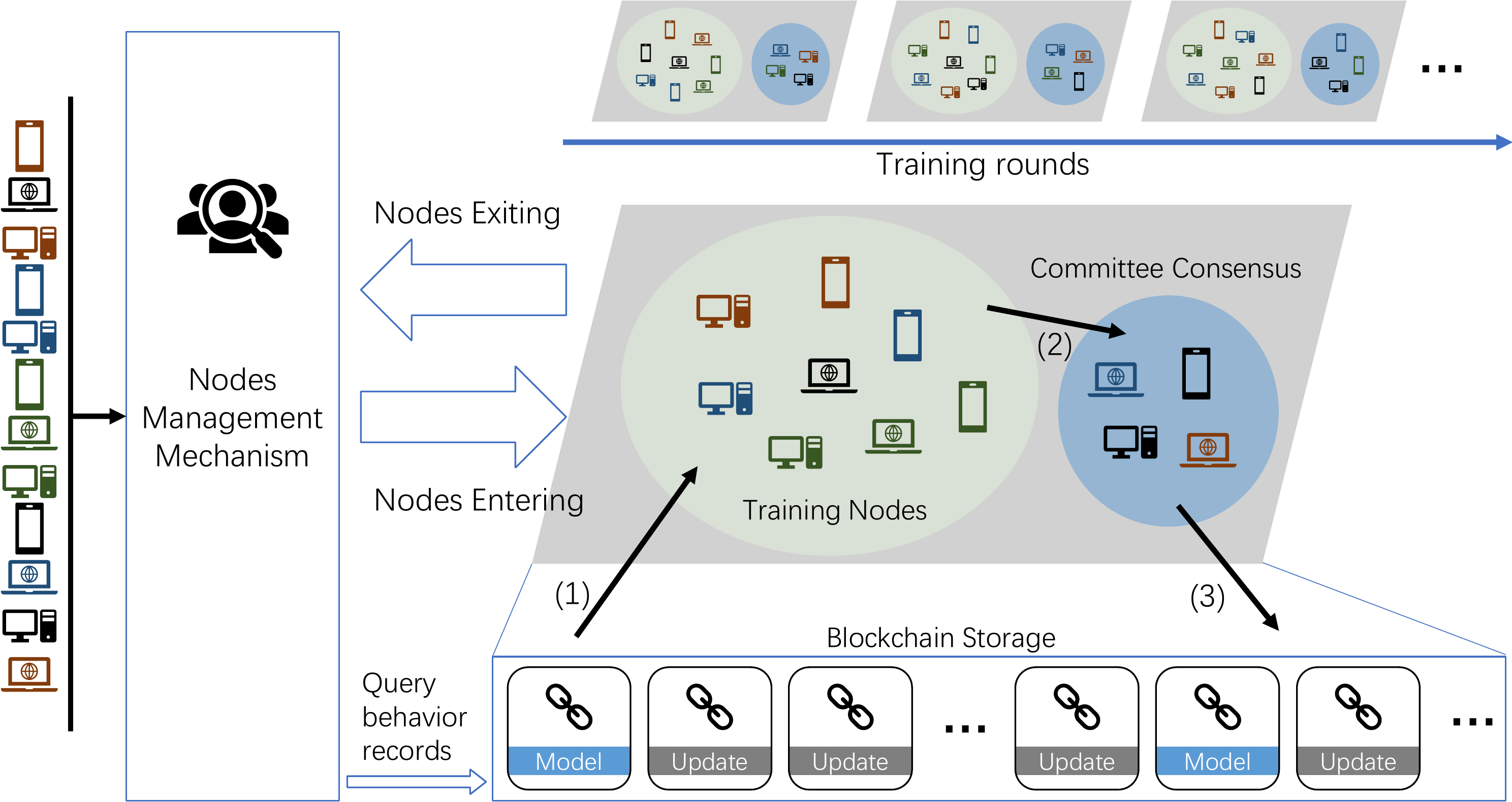}
    \caption{The training process of the proposed BFLC framework. (1) Training nodes acquire the newest global model and perform local training. (2) Training nodes send local updates to committee (3) Committee validate the updates and record new model or updates onto blockchain.
    }
    \label{fig:framework}
\end{figure*}

In this article, we propose a decentralized, autonomous blockchain-based FL architecture to address these challenges (shown in Figure \ref{fig:framework}). With respect to the management of FL nodes, the architecture based on the alliance chain ensures the node permission control. In terms of storage, we design the storage pattern on the chain of models and updates, and via this pattern, the nodes can get the latest model quickly. Each validated update is recorded and kept untampered on the blockchain. Considering the huge storage consumption on the blockchain, partial nodes can abandon the historical blocks to release the storage space. In terms of the block consensus mechanism, a novel \emph{committee consensus} mechanism is proposed, which can only increase a few validation consumptions and achieve more stability under malicious attacks. In each round of FL, updates are validated and packaged by a small number of nodes (i.e., the \emph{committee}). The committee consensus mechanism allows most honest nodes to reinforce each other and continuously improve the global model. A small number of incorrect or malicious node updates will be ignored to avoid damaging the global model. In the meantime, the BFLC training community is flexible, where the nodes can join or leave at any time without damaging the training process. Combined with an effective incentive mechanism, the nodes who contribute can gain actual rewards, thus promoting the development of the whole training community in a virtuous circle.

Our contributions are summarized as follows.
\begin{itemize}
    \item We propose a blockchain-based FL framework BFLC, which defines the model storage patterns, the training process and a novel committee consensus in detail.
    \item We technically discuss the scalability of BFLC, including the node management in the community, the analysis of malicious node attacks, and the storage optimization.
    \item We demonstrate the effectiveness of BFLC by experiments on real-world FL dataset. We also verify the security of BFLC by simulating the malicious attacks.
\end{itemize}

\section{Related Work}
Konecný \etal proposed Federated Learning whose goal is to train a high-quality centralized model while training data remains distributed over a large number of clients \cite{DBLP:journals/corr/KonecnyMYRSB16}
. The network situation of FL is unreliable and relatively slow, and the clients are not always online. In these years, FL is applied in many scenarios like video analysis, information inspection, and classification, credit card fraud detection while keeping the personal data sensitivity safe. Besides, the theoretical studies of convergence, network latency or malicious attacks on FL are also active fields.

The centralized federated server has been challenged and questioned growly in these years
. It is a natural thought that keeping the concept of server at a minimum or even avoiding it completely. The study of \cite{DBLP:conf/dais/HegedusDJ19} assumed that the data remains at the edge devices, but it requires no aggregation server or any central component. 
Hu \etal \cite{DBLP:journals/corr/abs-1908-07782} proposed a segmented gossip approach, which makes full utilization of node-to-node bandwidth then can achieve a convergence efficiently. 

Meanwhile, decentralization may be the most direct way to avoid the above risks. Blockchain, a distributed ledger technique, can store the historical operations and keep it tamper-resistant. With the aim of the blockchain, collaborative machine learning methods can get rid of the centralized server and improve security. Blaz \etal \cite{DBLP:journals/sensors/PodgorelecTK20} proposed a machine learning-based method to fasten the transaction signing process while also including a personalized identification of anomalous transactions. Deep reinforcement learning is also applied on blockchain-based scenarios, such as industrial Internet of things
, mobile edge computing
, cognitive radio networks
and Internet of vehicle
. 

In recent years, FL is an emerging research focus on the blockchain system.
It is reasonable to assume that the clients in FL might be malicious. Therefore, the local updates from all clients should be recorded under blockchain-based FL settings. You \etal \cite{DBLP:conf/apnoms/KimH19a} focused on the stability and convergence speed of FL, and proposed a blockchain-based method to address these challenges. Umer \etal \cite{DBLP:conf/apnoms/MajeedH19} propose a blockchain-based architecture, which can perform parallel learning for multiple global models. Bao \etal \cite{DBLP:conf/bigcom/BaoSXHH19} proposed a public blockchain-based FL architecture, which provides trusty consensus basing on nodes' data amount and historical performance.

These blockchain-based learning methods can effectively record the nodes' performance to reduce malicious attacks. However, there are still three main challenges:
\begin{enumerate}
    \item \emph{Consensus efficiency}. It is an inevitable process for blockchain-based methods to reach a consensus for each packing block. Considering the vast amount of learning nodes in the FL settings, a broadcasting consensus is highly time-consuming. Therefore, reducing the consensus cost is non-trivial. One \cite{DBLP:conf/bigcom/BaoSXHH19} of the related works selects a leader to execute the consensus. However, the criterion relies on many outer data.
    \item \emph{Model security}. The framework should prevent the global model from exposing to unauthorized devices and from poisoning. The security of the system is rarely studied under blockchain-based FL settings.
    \item \emph{Framework scalability}. When applying these training frameworks to real-world applications, we always need to add detail rules to adapt to different scenarios. Therefore, the scalability of frameworks determines their scope of applications.
\end{enumerate}
In the following sections, we will describe our proposed methods to tackle these challenges. 

\section{The Proposed Framework}

Federated Learning (FL) enables the machine learning algorithms training across multiple distributed clients without exchanging their data samples. In the original FL settings, one centralized server takes control of the training process, including client management, global model maintenance, and gradient aggregation. During each training round, the server broadcasts the current model to some participating nodes. After receiving the model, nodes locally update it with their local data and submit the update gradients to the server. The server then aggregates and applies the local gradients into the model for the next round.

The decentralized nature of blockchain can replace the place of the central server. As aforementioned, the functions of the centralized server can be implemented by the Smart Contract (SC) instead, and be actuated by transactions on the blockchain. To tackle this vision, we propose BFLC, which is a \underline{B}lockchain-based \underline{F}ederated \underline{L}earning framework with \underline{C}ommittee consensus. Without any centralized server, the participating nodes perform FL via blockchain, which maintains the global models and local updates. Considering the communication cost of FL, we leverage a novel delegated consensus mechanism to tackle the missions of gradients selection and blocks generation. In the following sub-sections, we will elaborate on the various components of the framework.

\subsection{Blockchain Storage}
\begin{figure}[t]
    \centering
    \includegraphics[width=.95\linewidth]{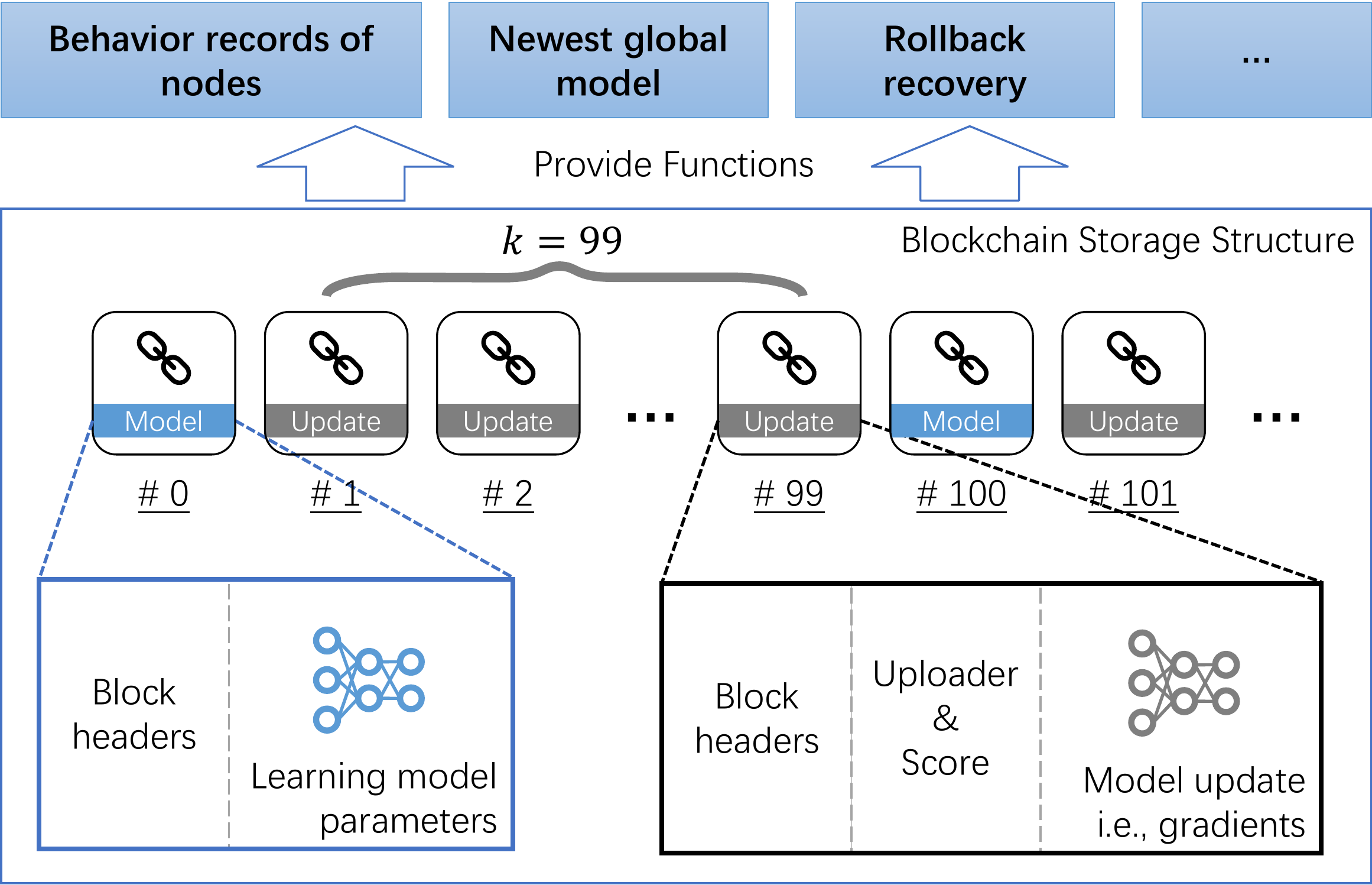}
    \caption{The FL storage structure on blockchain system, and the provided functions.}
    \label{fig:blockchain}
\end{figure}

To enable authority control, the storage of BFLC is an alliance blockchain system, and only the authorized devices can access the FL training contents. On the blockchain, we design two different blocks to store the global model and local update (as shown in Figure \ref{fig:blockchain}), which are collectively known as learning information. For the sake of simplicity, we assume that only one learning information is placed in a block.

In the beginning, a randomly initialized model was placed into the \#0 block, then the $0$-th round of training starts. Nodes access the current model and execute local training, and put the verified local gradients to new update blocks. When there are continuously enough update blocks, the smart contract triggers the aggregation, and a new model of the next round is generated and placed on the chain. We should note that the FL training only relies on the latest model block, and the historical block is stored for failure fallback and block verification.

We denote the number of required updates for each round as $k$, and denote the number of rounds as $t = 0, 1, ...$. Then we have: the \# $t \times (k+1)$ block contains the model of $t$-th round, which is called \emph{model block}, and the \# $[t \times (k+1)+1, (t+1) \times (k+1)-1]$ blocks contain the updates of $t$-th rounds, which are called \emph{update block}s. From an implementation perspective, one model block should includes: block headers, number of round $t$ and global model, while one update block includes: block headers, number of round $t$, local update gradient, uploader address and update score.

\subsection{Committee Consensus Mechanism}
The chain structure of blockchain guarantees the immutability. Therefore, appending the correct blocks to the chain is a crucial component which is the consensus mechanisms work for. The competition-based consensus mechanisms append blocks on the chain first, whereafter, the consensus meets. Conversely, the communication-based generate mechanisms reach an agreement before appending blocks.

Considering the computation and communication cost of consensus, we propose an efficient and secure Committee Consensus Mechanism (CCM) to validate the local gradients \emph{before} appending it to the chain. Under this setting, a few honest nodes will constitute a \emph{committee} in charge of verification of local gradients and blocks generation. In the meantime, the rest nodes execute local training and send the local updates to the committee. The committee then validates the updates and assign a score on them. Only the qualified updates will be packed onto the blockchain. At the beginning of the next round, a new committee is elected basing on the scores of nodes in the previous round, which means that the committee will not be re-elected. It is noteworthy that the update validation is a pivotal component of the CCM, therefore, we describe a feasible approach: the committee members validate the local updates by treating their data as a validation set, and the validation accuracy becomes the score. This is the minimized approach that acquires no further operation of the committee, but only the basic ability to run the learning model. After combining the scores from the various committee members, the median will become the score of this update.

Working with this mechanism, BFLC can achieve these advantages:
\begin{enumerate}
    \item \emph{High efficiency}: only a few nodes will validate the updates, rather than broadcasting to every node and reach an agreement.
    \item \emph{K-fold cross-validation}: the committee members will not participate in the local training in the round. Therefore, the local data of the committee are taken as a validation set. As the alternating of committee members at each round, the validation set changes as well. In this setting, k-fold cross-validation on FL achieved.
    \item \emph{Anti-malevolence}: based on the validation scores, the corresponding nodes with better performance will be elected by the smart contract and constitute the new committee for the next training round. which means the selected local data distribution is gregarious and the node is not malicious.
\end{enumerate}

\subsection{Model Training}
Nodes other than committees perform local training each round. In FL, for the sake of security and privacy, raw data will be kept in nodes locally, and these nodes only upload the gradients to the blockchain. Furthermore, there are two main challenges:
\begin{enumerate}
    \item The local data distribution might be not Independent and Identically Distributed (non-IID)
    \item The devices are not always available
\end{enumerate}
To address the first challenge, only a certain number of local updates are requisite for each round, and the committee consensus mechanism could maximize the generalization ability of the global model by validating the local updates with committee members' data distribution \cite{DBLP:journals/corr/KonecnyMYRSB16}. To address the second one, we design an initiative local learning progress for nodes.

Nodes can actively obtain the current global model at any time and perform local training. The gradients will be sent to the committee and be validated. When eligible updates are packaged on the blockchain, as a reward, tokens can be attached to them. We will discuss the incentive in the next section.

As aforementioned, a certain number of valid updates are required for each round. Therefore, when the committee validates enough local updates, the aggregation process is activated. These validated updates are aggregated by the committee into a new global model. The aggregation can be performed on the local gradients \cite{DBLP:conf/ccs/ShokriS15} or the local models \cite{DBLP:conf/aistats/McMahanMRHA17}, and the network transmission consumptions of these two methods are equal.
After the new global model is packed on the blockchain, the committee will be elected again, and the next training round begins.

\section{Discussion}
\subsection{Node Management and Incentive}
The BFLC training process depends on the mutual promotion of nodes, and node management is also a key part of BFLC. The participant nodes can not only access the global model but can also upload updates to affect the global model. To control permissions, we have designated the initial nodes that constitute the training community to be responsible for node management, i.e., to be the managers. Each device must be verified by the managers before joining the training community. This verification is in blacklist mode: if the device has been kicked out of the community for misconduct (such as submitting misleading updates, spreading a private model), the device will be rejected.

Depending on the proposed blockchain storage structure, the latest global model can be quickly found on the chain after new nodes joined. Nodes can immediately use the model to complete their local tasks, or they can update the model with local data and gain scores on the chain after verification by the consensus committee. It is noteworthy that only a certain number of valid updates are required for aggregation at each round, and only part of nodes are online to participate as well. Therefore, as long as the nodes actively submit updates, it is likely to participate in the global model updating and gain scores. Meanwhile, partial offline nodes will not impede FL progress.

Nodes in a community can always use the model without committing updates, so an effective incentive is required to encourage nodes to provide updates to the global model. To address this problem, we propose an incentive mechanism called \emph{profit sharing by contribution}.
\begin{itemize}
    \item \emph{Permission fee}: each device should pay for the access permission of the global model, and these fees are kept by the managers. Nodes then have unlimited access to the latest models in the community.
    \item \emph{Profit sharing}: After aggregation of each round, the managers distribute rewards to the corresponding nodes basing on the scores of their submitted updates.
\end{itemize}
As a result, frequently providing updates could earn more rewards, and the constantly updated global model will attract more nodes to participate. This incentive mechanism has high scalability to adapt to different real-world applications. For instance, the configurations of the permission fee, the profit-sharing proportion or the dividend modes are worth studying.

\subsection{Committee Election}
At the end of each round, a new committee is elected from the providers of validated updates. In decentralized training settings, this election significantly affects the performance of the global model, because the committee decides which local updates will be aggregated. Committee election methods include the following categories:
\begin{itemize}
    \item \emph{Random election}: new committee members are randomly selected from validated nodes. From a machine learning perspective, this approach improves the generalization of the model and reduces overfitting. However, the resistance to malicious attacks is weak while the malicious nodes disguise as normal ones.
    \item \emph{Election by score}: the providers with top validation scores constitute the new committee. This may exacerbate the uneven distribution of samples due to the absence of partial nodes into the committee. However, for malicious node attacks, this approach significantly increases the cost of the attack and brings more security and stability.
    \item \emph{Multi-factor optimization}: this approach considers multiple factors of the device (i.e., the network transmission rate) and the validation scores for optimal election. However, this optimization will bring additional computing overhead. Therefore, this approach should be applied depending on the realistic scenario and the associated requirements.
\end{itemize}

\subsection{Malicious Nodes}
A malicious node is defined as a node submitting incorrect, malicious model updates. In the original FL settings, FedAvg \cite{DBLP:conf/aistats/McMahanMRHA17} aggregates all the updates into a new global model. If there are malicious updates, the global model will be poisoned and obtains lower performance. As aforementioned, under the CCM, the updates will be verified by the committee before being aggregated. In this sub-section, we theoretically analyze the factors and the success possibility of malicious attacks.

We denote the amount of all nodes as $N$, in which the amount of committee members is $M$, and the rest $N-M$ nodes are training nodes. Distinctly, a malicious update is accepted to the aggregation if and only if more than $\frac{M}{2}$ committee members are cooperating with. However, the committee members are the $M$ of the best performers at last round, which means these malicious committee members' updates are accepted by other $\frac{M}{2}$ malicious nodes in the last committee. It is an infinite dependency loop, therefore, as long as there are more than $\frac{M}{2}$ honest nodes in the first committee, no malicious node could enter the committee and harm the global model.

\begin{figure}[t]
    \centering
    \includegraphics[width=.7\linewidth]{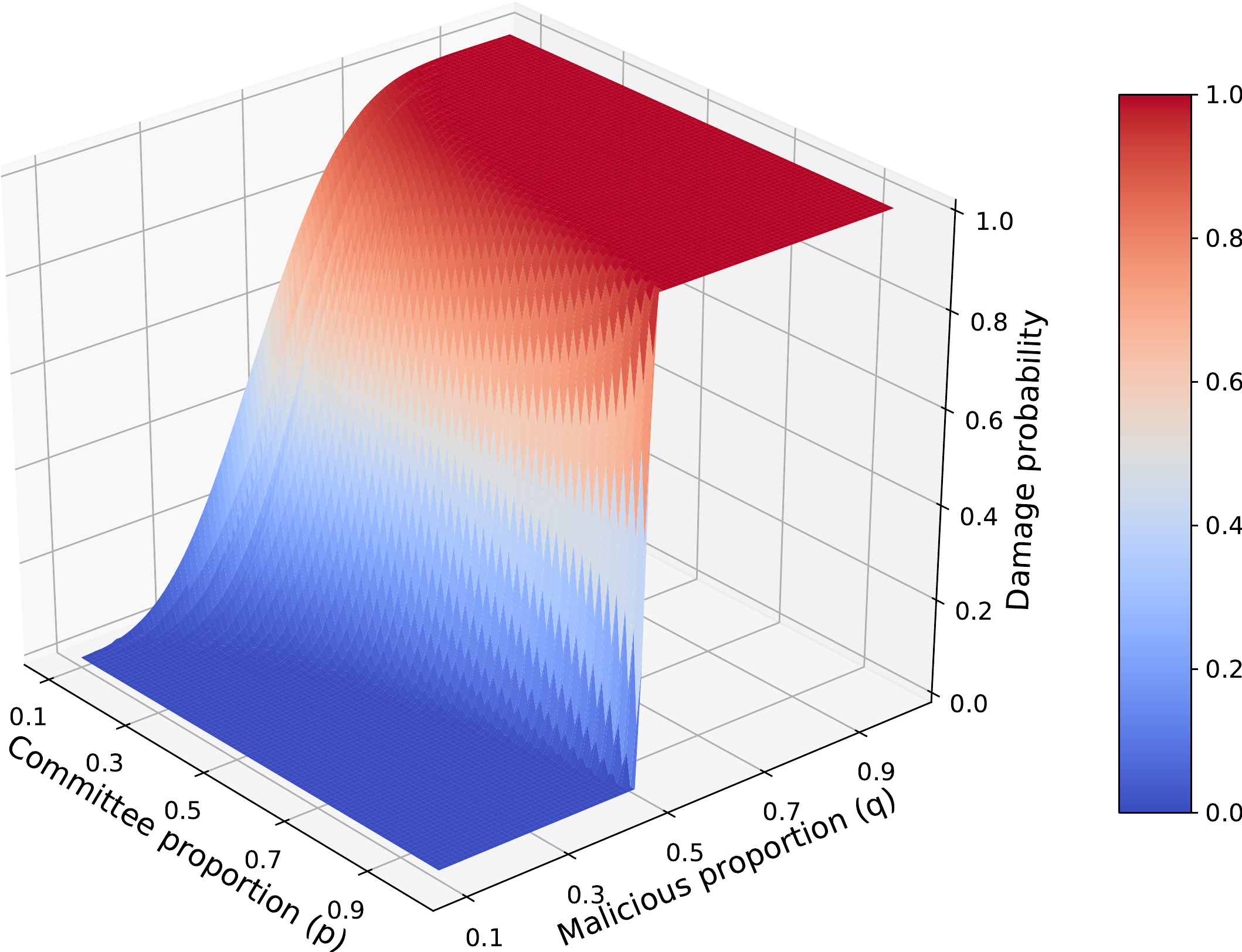}
    \caption{The attack success probability changing along with $p$ and $q$.}
    \label{fig:out}
\end{figure}

Considering another extreme situation: the malicious nodes conspire together to earn the committee seats by pretending as normal nodes. When the malicious nodes hold half of the seats, the attack begins. To analyze this attack mode, we denote the amount of participating nodes as $A$, the percentage of malicious nodes in $A$ is $q \in (0, 1)$, and the percentage of the committee is $p \in (0, 1)$. The attack target is holding more than $\frac{A \times p}{2}$ seats in committee. We assume that the performance of each node is similar. Therefore, the attack success probability can be calculated as the probability of this event: extracting $A \times p$ nodes from $A$ nodes, more than half of which come from $A \times q$.
By fixing $A=1000$, we plot the probability change along $p$ and $q$ in Figure \ref{fig:out}. We should note that, only when the malicious percentage greater than 50\%, the attack success probability could be greater than 0 markedly. This conclusion is similar to the 51\% attack in the PoW blockchain system. In other words, in a decentralized community, the malicious nodes should hold 51\% computational resources to attack the system, where the cost far outweighs the benefit. Furthermore, the historical models and updates are stored on the blockchain, therefore, failback is always an option after the attack happened.

\begin{table*}[!t]
    \centering
    \caption{Accuracy of BFLC, Basic FL and stand-alone on FEMINST dataset with different proportion of active nodes }
    \label{tab:feminst}
    \begin{tabular}{c|ccccc} 
    \hline
    \multirow{2}{*}{Frameworks} & \multicolumn{5}{c}{proportion $k$\% of active nodes}  \\ 
    \cline{2-6}
                                & 10\%    & 20\%    & 30\%    & 40\%    & 50\%          \\ 
    \hline
    BFLC                        & 89.33\% & 89.89\% & 90.02\% & 89.87\% & 89.78\%       \\ 
    \hline
    Basic FL                    & 90.02\% & 90.20\% & 90.29\% & 90.11\% & 90.42\%       \\ 
    \hline
    Stand-alone                 & \multicolumn{5}{c}{91.34\%}                                 \\
    \hline
    \end{tabular}
\end{table*}

\subsection{Storage Optimization}
In real-world applications, storage overhead is an important factor that determines the hardware requirements for the training devices. Based on the above-mentioned blockchain storage scheme (as shown in Figure \ref{fig:blockchain}), the latest global model can be found quickly. Although historical models and updates can provide post-disaster recovery functions, they also occupy huge storage space. Here, we give a simple and feasible storage overhead reduction scheme: nodes with insufficient capacity can delete historical blocks locally, and only keep the latest model and updates of the current round. In this way, the problem of insufficient storage space on some nodes can be solved, while the ability to recover from disasters and block verification is retained on the core nodes. However, the shortcomings of this method are also obvious. The credibility of the blockchain decreases with the deletion of nodes. In a mutually distrusting training community, each node may not use this scheme for security concerns.

Therefore, trusted and reliable third-party storage may be a better solution. The blockchain only maintains the network address where each model or updated file is located and records of modification operations. Other nodes interact with the centralized storage to obtain the latest model or upload updates. This centralized storage will be responsible for disaster recovery backup and distributed file storage services.

\section{Experimental}
\subsection{Settings and Normal Training}
To demonstrate the effectiveness of the BFLC, we perform it on the real-world federated dataset FEMNIST \cite{DBLP:journals/corr/abs-1812-01097}. This dataset contains 80,5263 samples and 3550 users for handwritten character image classification tasks and contains 62 different classes (10 digits, 26 lowercase, 26 uppercase). Following the instruction of the dataset, we simulate 900 devices in the training community, where the local datasets are unbalanced in number and not independent in distribution. After the active nodes are randomly selected, we perform the local training and aggregate via memory. We employed a blockchain system named FISCO \footnote{https://github.com/fisco-bcos} with PBFT consensus on an Intel Core CPU i9-9900X with a clock rate of 3.50 GHz with 10 cores and 2 threads per core. The SC layer was developed using the Solidity programming language. The learning model is written with Python 3.7.6 and Tensorflow 1.14.0 and is executed on Geforce RTX 2080Ti GPUs.

We compare the BFLC with the basic FL \cite{DBLP:conf/aistats/McMahanMRHA17} framework and the stand-alone training framework as the baseline. Each framework performs the classic image classification model AlexNet \cite{DBLP:journals/cacm/KrizhevskySH17} as the global model and fixed a set of model hyper-parameters to ensure fairness. In terms of the experimental settings, we defined the proportion of active nodes in each round as $k$\%, among which 40\% will be elected as committee members in the next round for BFLC. The proportion of training nodes for Basic FL is also $k$\%. Meanwhile, stand-alone training will leverage the whole dataset. Under the conditions of different $k$ values, we recorded their performance in Table \ref{tab:feminst}.

As can be seen in Table \ref{tab:feminst}, with the increase in the proportion of active nodes, the performance of BFLC keeps approaching the effect of the basic FL framework and only has a slight loss compared to the stand-alone training with the intact dataset. We should mention that the BFLC can significantly reduce the consumption of consensus through the committee consensus mechanism. For instance, if the number of training nodes is $P$ and the size of the committee is $Q$, then the active nodes are $P+Q$. For BFLC, the amount of calculation per round of consensus can be expressed as $P \times Q$, while the broadcasting approach is $(P+Q)^2$. Compared with stand-alone training, BFLC also has the privacy data protection function of federated learning and does not require a trusted central server to manage, which significantly reduces the risk of privacy leakage.

\subsection{Under Malicious Attack}
The malicious nodes in the training community will generate harmful updates, which will significantly reduce the performance of the global model if being integrated. In this sub-section, we simulate malicious node attacks to demonstrate how the proposed BFLC, basic FL, and CwMed \cite{DBLP:conf/icml/YinCRB18} will be affected under different malicious proportions among active nodes. We assume that the attack mode of the malicious node is random perturbation with a pointwise Gaussian random noise.

The basic FL will not perform any defense measures, and model updates generated by randomly selected active nodes will be integrated. CwMed constructs a global gradient, where each entry is the median of entries in the local gradients with the same coordinate. BFLC relies on the committee consensus mentioned above to resist the attack. Each update will obtain a score from the committee (i.e., the median local prediction accuracy).

\begin{figure}[t]
    \centering
    \includegraphics[width=.7\linewidth]{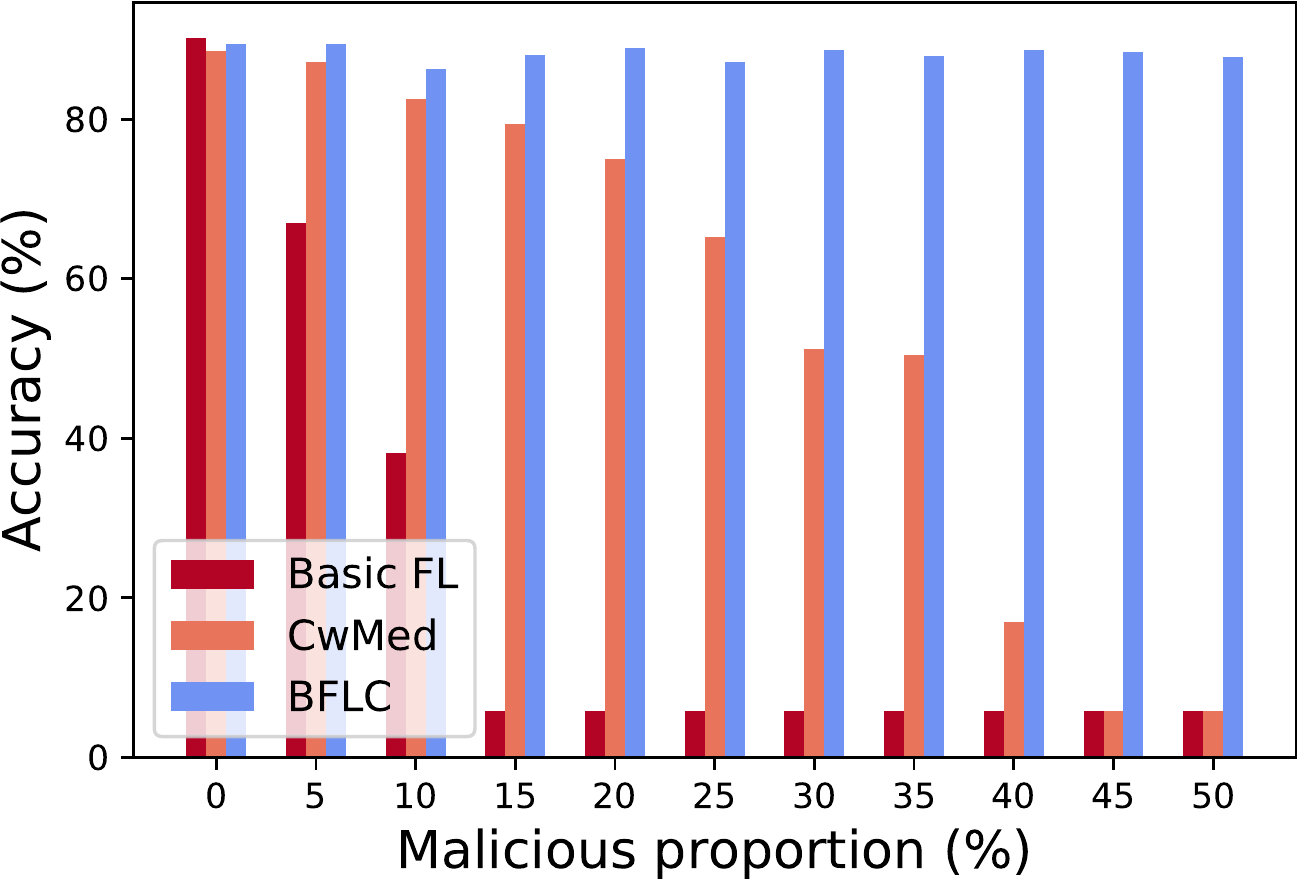}
    \caption{Performance of methods under malicious attacks.}
    \label{fig:out2}
\end{figure}

In order to enhance the effectiveness of the attack, we assume that malicious nodes are collusion, that is, members of the malicious committee will give random high scores (for example, 90\% \~ 100\%) to the malicious updates. The proportion of active nodes is fixed as 10\%, and 20\% of them will be elected as the committee in the next round. As shown in Figure \ref{fig:out2}, the BFLC can resist much higher malicious nodes proportion than the compared methods. This indicates the effectiveness of BFLC with the help of the committee mechanism.

\section{Conclusion}
The security of federated learning is facing increasing challenges. Malicious FL training nodes would damage the global model, while a malicious central server can also leak the nodes' private data. Based on a trusted blockchain system, we propose BFLC, which is a decentralized, federated learning framework exploiting the committee consensus. Such the committee consensus can effectively avoid the influence of malicious central servers or malicious nodes. In the experiment section, we verified the effectiveness of the BFLC framework by adopting the real-world dataset, which can obtain a global model similar to the centralized training in a federated learning framework. We also discussed the scalability of BFLC, which has broad research prospects in security, data storage, and incentive mechanisms.

\section{Acknowledgement}
The work described in this paper was supported by the National Key Research and Development Program (2016YFB1000101), the National Natural Science Foundation of China (11801595, 61722214), the Natural Science Foundation of Guangdong (2018A030310076, 2019A1515011043) and the CCF-Tencent Open Fund WeBank Special Funding.

\bibliographystyle{IEEEtran}
\bibliography{bflc}

\begin{IEEEbiographynophoto}
    {Yuzheng~Li}
    received the B.S. degrees from the Sun Yat-sen University, Guangdong, China, in 2018. He is currently pursuing an M.S. degree with the School of Data and Computer Science, Sun Yat-sen University. His current research interests include federated learning, blockchain, statistical machine learning, multi-view learning, and optimization.
\end{IEEEbiographynophoto}

\begin{IEEEbiographynophoto}
    {Chuan~Chen}
    received the B.S. degree from Sun Yat-sen University, Guangzhou, China, in 2012, and the Ph.D. degree from Hong Kong Baptist University, Hong Kong, in 2016. He is currently an Associate Research Fellow with the School of Data and Computer Science, Sun Yat-sen University. His current research interests include blockchain, machine learning, numerical linear algebra, and numerical optimization.
\end{IEEEbiographynophoto}

\begin{IEEEbiographynophoto}
    {Nan~Liu}
    received the B.S. degree from the Sun Yat-sen University, Guangzhou, China, in 2019. He is currently pursuing an M.S. degree with the School of Data and Computer Science, Sun Yat-sen University, Guangzhou, China. His research interests include federated learning, blockchain, and network representation learning.
\end{IEEEbiographynophoto}

\begin{IEEEbiographynophoto}
    {Huawei~Huang}
    received his Ph.D. in Computer Science and Engineering from the University of Aizu, Japan. He is currently an associate professor with the School of Data and Computer Science, Sun Yat-Sen University, China. His research interests mainly include distributed learning and blockchain. He has served as a visiting scholar with the Hong Kong Polytechnic University (2017-2018); a post-doctoral research fellow of JSPS (2016-2018); and an assistant professor with Kyoto University, Japan (2018-2019).
\end{IEEEbiographynophoto}

\begin{IEEEbiographynophoto}
    {Zibing~Zheng}
    received his PhD degree from the Chinese University of Hong Kong in 2011. He is currently a professor in the School of Data and Computer Science at Sun Yat-sen University, China. He has published over 120 international journal and conference papers, including three ESI highly cited papers. According to Google Scholar, his papers have more than 9,100 citations, with an II-index of 46. His research interests include blockchain. services computing, software engi­neering, and financial big data.
\end{IEEEbiographynophoto}

\begin{IEEEbiographynophoto}
    {Qiang~Yan}
    is currently the Blockchain Scientist at WeBank. He received his Ph.D. degree in Information Systems from Singapore Management University in 2013. Before joining WeBank, he was the tech lead of the privacy infrastructure team at Google Switzerland. His research interests include applied security and privacy for blockchain, AI, mobile systems, social networks and more, human factors in security system design, and applied cryptography.
\end{IEEEbiographynophoto}
\end{document}